\documentclass[aps,prc,twocolumn,nofootinbib]{revtex4-1}

\usepackage{xspace}
\usepackage{graphicx}
\usepackage{amsmath}
\usepackage[]{color}
\usepackage[bookmarks=false,colorlinks=true,citecolor=red,urlcolor=cyan,%
   pdfstartview=FitH]{hyperref}
\usepackage{txfonts}

\newcommand{\nuc}[2]{${}^{#1}$#2}
\renewcommand{\today}{\number\day\space\ifcase\month\or January
 \or February\or March\or April\or May\or June\or July\or August
 \or September\or October\or November\or December\fi\space\number\year}
\newcommand{\degreesC}{$\,^{\circ}$C\xspace}
\newcommand{\jour}[4]{{#1} {\bf #2}, {#3} ({#4})}

\begin{document}

\title{
       Gallium experiments with artificial neutrino sources \\
       as a tool for investigation of transition to sterile states
      }
\author{V.\,N.\,Gavrin}
\thanks{
        Institute for Nuclear Research,
        Russian Academy of Sciences, 117312 Moscow, Russia.
       }
\author{V.\,V.\,Gorbachev}
\thanks{
        Institute for Nuclear Research,
        Russian Academy of Sciences, 117312 Moscow, Russia.
       }
\author{E.\,P.\,Veretenkin}
\thanks{
        Institute for Nuclear Research,
        Russian Academy of Sciences, 117312 Moscow, Russia.
       }
\author{B.\,T.\,Cleveland}
\thanks{University of Washington, Seattle, Washington 98195, USA.}

\affiliation{\vspace{-2ex}}
\author{\vspace{-2ex}20 July 2010}

\noaffiliation

\begin{abstract}

We propose to place a very intense source of \nuc{51}{Cr} at the
center of a 50--tonne target of gallium metal that is divided into
two concentric spherical zones and to measure the neutrino capture
rate in each zone.  This experiment can set limits on transitions
from active to sterile neutrinos with $\Delta m^2 \approx 1$~eV$^2$
with a sensitivity to disappearance of electron neutrinos of a few
percent.

\end{abstract}


\maketitle

  \pagestyle{myheadings}
  \markboth{\hfill 20 July 2010 \hfill 2-zone Ga source experiment}
           {2-zone Ga source experiment \hfill 20 July 2010 \hfill}
  \topmargin=-15mm 
  \headheight=12pt
  \headsep=25pt         


Analysis of neutrino experiments has given convincing evidence of
transitions between flavors, i.e., neutrino oscillations.  These
neutrino transitions are well described in the framework of three
neutrino generations with masses $m_1, m_2, \text{ and } m_3$ whose
mass-squared differences are 
$\Delta m_{12}^2 \approx 8 \times 10^{-5}$~eV$^2$ (``solar'') and 
$\Delta m_{23}^2 \approx 2 \times 10^{-3}$~eV$^2$ (``atmospheric'').
Almost all neutrino experiments can be explained by assuming that
there are only these three neutrino generations.

The existence of three dominant neutrino generations has been further
proven by experiments at LEP on the decay of the $Z_0$ boson.  The
LEP experiment, however, does not rule out the possibility that there
may be additional sub-dominant neutrino species and there are some
indications that the number of neutrino generations may be more than
three.  First and foremost is the accelerator experiment LSND whose
results can be explained if there are neutrino transitions with
$\Delta m^2 \approx 1$~eV$^2$.  Such a large value cannot be obtained
with three neutrino generations as then the maximal value is $\Delta
m^2 \approx \Delta m_{23}^2 \approx 2 \times 10^{-3}$~eV$^2$.  To
comply with the results of LEP experiments one must assume that the
hypothesized fourth state of neutrino flavor has an interaction cross
section with matter that is much less than the other three neutrino
species, i.e., it must be ``sterile''.

The idea of the sterile neutrino was first proposed by B. Pontecorvo
\cite{Pontecorvo67} and has been used repeatedly to explain a variety
of different neutrino observations \cite{Kusenko09}.  For example, to
explain the slowly-rising spectra of recoil electrons from \nuc{8}{B}
solar neutrinos in the region below 6--8 MeV in the experiments
SuperK and SNO, Smirnov and Holanda \cite{smirnov04} considered
transitions to sterile states with $\Delta m^2 \approx 10^{-5}$
eV$^2$.
Proposals for several new experiments to search for sterile neutrinos
are in Ref.~\cite{other_sterile_exp}.

Another set of experiments that can be interpreted in terms of
neutrino oscillations is the capture rate measurements in the Ga
detectors SAGE \cite{CrPRC,ArPRC} and GALLEX \cite{KaetherPLB}.
These four experiments used very-intense reactor-produced
\nuc{51}{Cr} or \nuc{37}{Ar} neutrino sources and their
weighted-average value, expressed as the ratio $R$ of the measured
production rate to the expected production rate based on the measured
source strength, is $0.87 \pm 0.05$, considerably less than the
expected value of unity.  All possible explanations for this
unexpectedly low result are discussed in detail in Ref.~\cite{art08}.
Foremost among these is overestimation of the cross section for
neutrino capture to the lowest two excited states in \nuc{71}{Ge},
which could yield a value of $R$ as small as 0.95.  Other
explanations might be a statistical fluctuation or a real physical
effect of unknown origin, such as a transition to sterile neutrinos
or quantum decoherence in neutrino oscillations \cite{farzan08}.

\begin{figure}
\centering
\includegraphics[width=0.8\hsize,viewport=17 20 278 215]{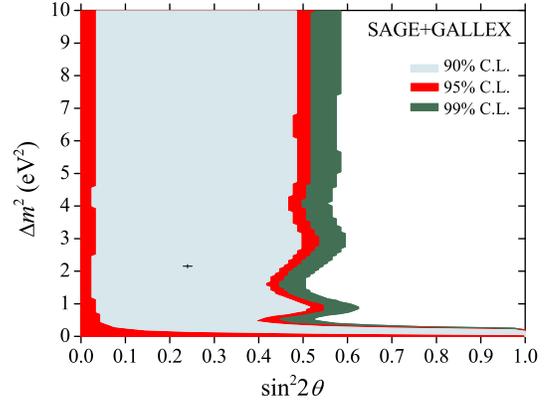}

\caption{Region of allowed mixing parameters inferred from gallium
source experiments assuming oscillations to a sterile neutrino.  Plus
sign at $\Delta m^2=2.15$~eV$^2$ and $\sin^2 2\theta=0.24$ indicates
the best-fit point.}

\label{sage-gallex}
\end{figure}

The interpretation of the Ga source experiments in terms of
oscillations to a sterile neutrino with $\Delta m^2 \approx
1$~eV$^2$, as well as the agreement of these results with the reactor
experiments Bugey, Chooz, and G\"osgen and the accelerator
experiments LSND and MiniBooNE is considered in detail in
Ref.~\cite{giunti}.  If transitions to a sterile neutrino are
occurring, the region of allowed oscillation parameters inferred from
the four Ga source experiments is shown in Fig.~\ref{sage-gallex}.

\begin{figure}
\centering
\includegraphics[width=0.6\hsize,viewport=29 16 185 194]{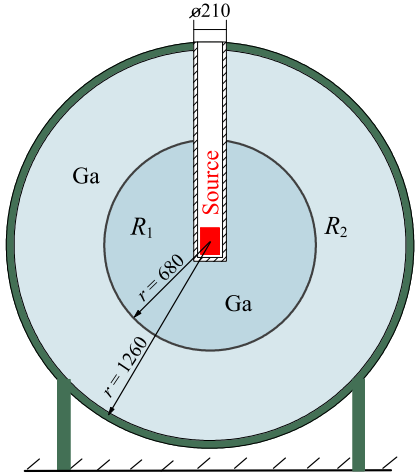}

\caption{Schematic drawing of proposed neutrino source experiment.
$R_1$ and $R_2$ are the ratios of measured capture rate to predicted
rate in the inner and outer zones, respectively.  Outer radii $r$ of
the two zones and diameter $\phi$ of source reentrant tube are
indicated in mm.}

\label{src_scheme}
\end{figure}

We believe that new experiments are necessary to understand the low
result of the Ga source measurements.  One such experiment is now in
progress at the RCNP cyclotron that should provide information to
better determine the cross section for neutrino capture at low energy
\cite{Ejiri}.  Another experiment, which we intend to pursue, is an
improved version of the Ga source measurements.  Our plan, as
schematically pictured in Fig.~\ref{src_scheme}, is to place a source
of \nuc{51}{Cr} with initial activity of 2~MCi at the center of a
50-tonne target of liquid Ga metal that is divided into two
concentric spherical zones, an inner 8-tonne zone and an outer
42-tonne zone.  If the neutrino capture cross section is that
calculated by Bahcall \cite{gacross} and oscillations to sterile
neutrinos do not occur, then at the beginning of irradiation there is
a mean of 43~atoms of \nuc{71}{Ge} produced by the source per day in
each zone.  After an exposure period of a few days, the Ga in each
zone is transferred to reaction vessels and the \nuc{71}{Ge} atoms
produced by neutrino capture are extracted.  These atoms are placed
in proportional counters and their number determined by counting the
Auger electrons released in the transition back to \nuc{71}{Ga},
which occurs with a half life of 11.4~days.  A series of exposures is
made, each of a few days duration, with the \nuc{71}{Ge} atoms from
each zone measured in individual counters.

These procedures are all well understood and were used in the prior
source experiments \cite{CrPRC,ArPRC}.  A Monte Carlo simulation
based on a reasonable time sequence of extractions and using typical
values of extraction efficiency, counter efficiency, and background
rates indicates that the rate in each zone can be measured with a
total uncertainty, statistical plus systematic, of $\sim$5\%.

If oscillations to a sterile neutrino are occurring with mass-squared
difference of $\Delta m^2$ and mixing parameter $\sin^2 2\theta$ then
the rates in the outer and inner zones of gallium will be different
and their ratio, for the specific case of $\sin^2 2\theta = 0.3$,
will be as shown in Fig.~\ref{rate_ratio}.  To see how this new
2-zone experiment will aid in the interpretation of the Ga source
measurements and may shed light on transitions to sterile neutrinos,
let us consider several possible outcomes and the inferences
therefrom:

\begin{figure}
\centering
\includegraphics[width=0.8\hsize,viewport=17 20 278 215]{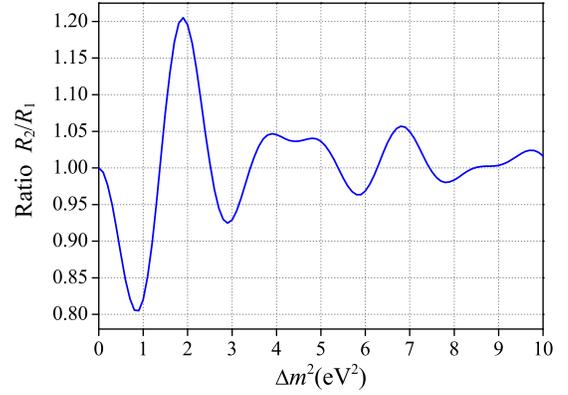}

\caption{Ratio of rates in  the outer and inner zones versus
$\Delta m^2$ for the case of $\sin^2 2\theta = 0.3$.}

\label{rate_ratio}
\end{figure}

\begin{figure*}
\centering
\hspace{1em}\includegraphics*[width=0.4\hsize,viewport= 17 20 279 217]{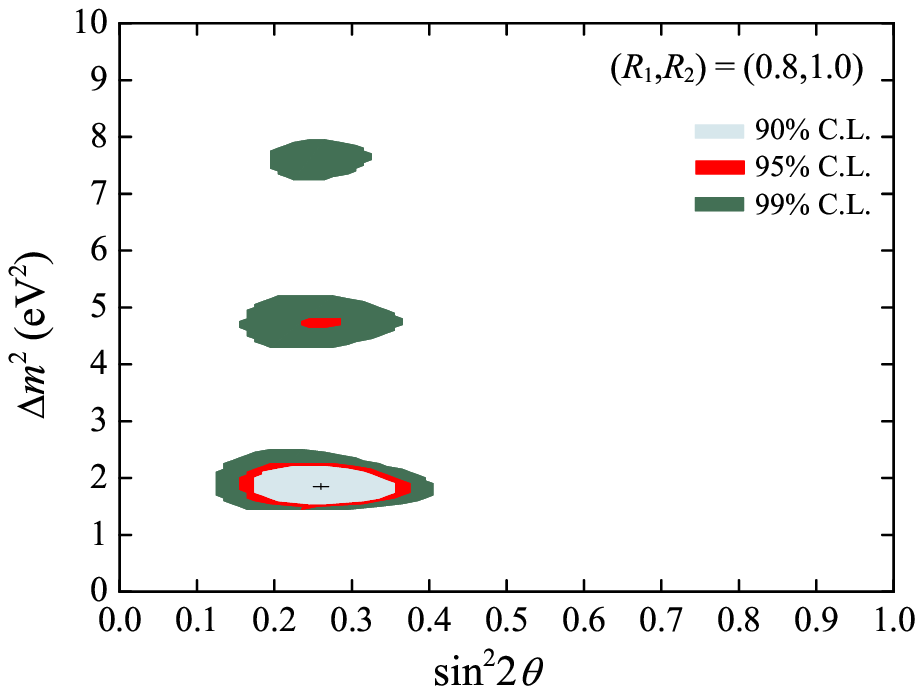}%
\hspace{1em}%
\includegraphics*[width=0.4\hsize,viewport= 17 20 279 217]{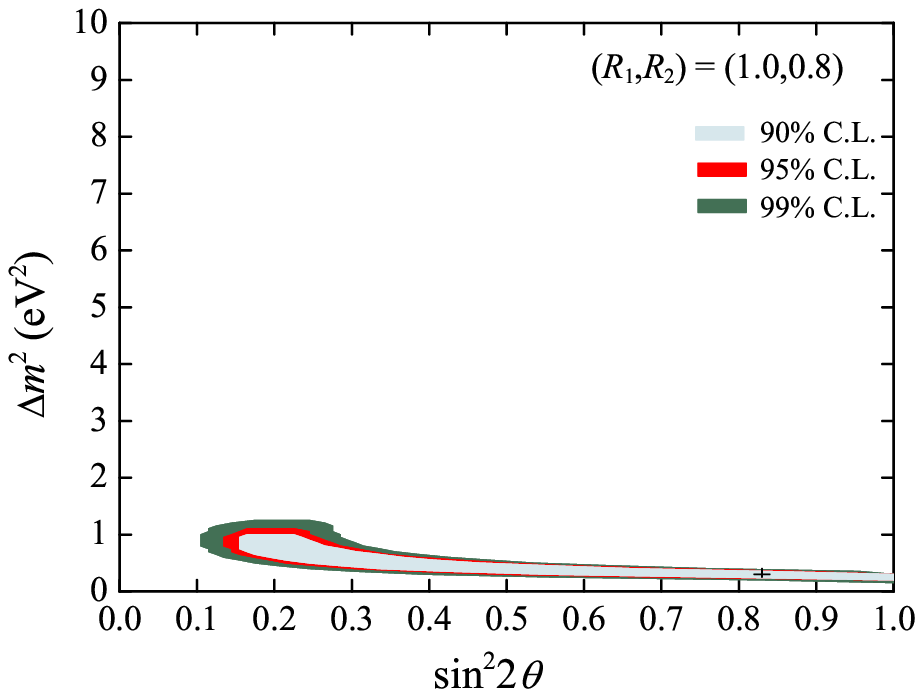}%
\vspace*{2em}
\hspace{1em}\includegraphics*[width=0.4\hsize,viewport= 17 20 279 217]{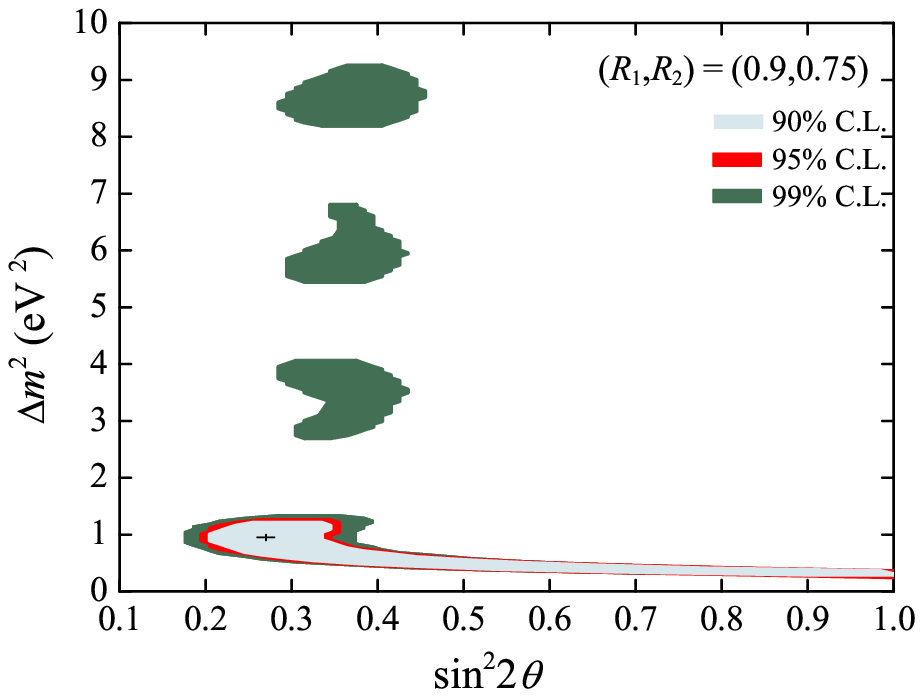}%
\hspace{1em}%
\includegraphics*[width=0.4\hsize,viewport= 17 20 279 217]{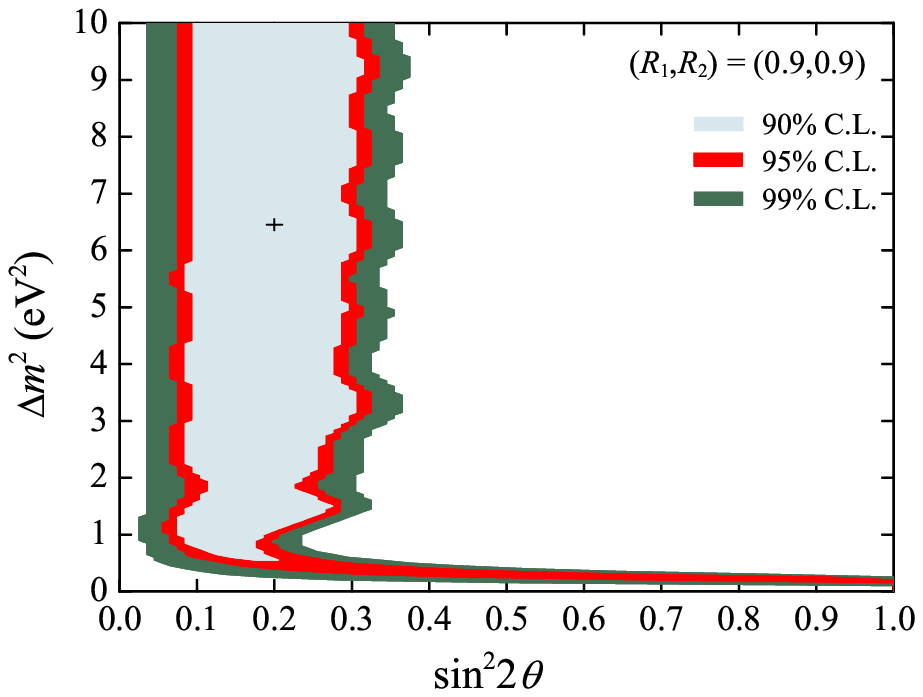}

\caption{Regions of allowed mixing parameters for various possible
outcomes of new two-zone Ga experiment.  $R_1$ and $R_2$ are the
ratios of measured rate to predicted rate in the inner and outer
zones, respectively.  Plus sign indicates the best-fit point.}

\label{regions}
\end{figure*}

\begin{itemize}

\item If the ratio of rates $R$ in the two zones are statistically
incompatible, such as $R_1=1.00 \pm 0.05$ in the inner zone and
$R_2=0.80 \pm 0.04$ in the outer zone, or vice versa, then it is
likely that transitions to sterile neutrinos are occurring.  We show
in the upper panels of Fig.~\ref{regions} the inferences that can be
drawn regarding the neutrino mixing parameters for these two possible
outcomes of the experiment.

\item If the ratio of rates $R$ in both zones are statistically
compatible, then the inferences will depend on what the average value
may be.  For example, if the average value is $0.95 \pm 0.034$ then
it is most likely that there is an error in the neutrino capture
cross section to low-energy states.  This inference is not clear-cut,
however, as average results in this range can also occur due to
transitions to sterile neutrinos.  In contrast, average values of $R$
much below 0.92 can only be due to transitions to sterile neutrinos.
As an example, the bottom right panel of Fig.~\ref{regions}
illustrates what can be learned regarding neutrino oscillation
parameters if the experimental result is $(R_1,R_2)=(0.90,0.90)$.

\end{itemize}

As is evident, it is only with certain specific outcomes that a
two-zone Ga experiment will unambiguously differentiate between an
oscillation interpretation or other possible interpretations of the
Ga source anomaly.

Nonetheless, our proposed Ga experiment has several significant
advantages over other methods of investigation of oscillations with
parameters $\Delta m^2 \approx 1$~eV$^2$:

\begin{itemize}

\item 90\% of the decays of the \nuc{51}{Cr} source give a neutrino
with energy 750~keV and 10\% an energy of 430~keV.  This
nearly-monochromatic energy is important for oscillation experiments
because the energy occurs in the denominator of the transition
probability $P$ in the form
\begin{equation}
P(\nu_e \rightarrow \nu_e) = 1 - \sin^2(2\theta) \sin^2[1.27\Delta m^2(\text{eV}^2) \frac{L}{E_{\nu}}]
\end{equation}
where $\theta$ is the mixing angle, $L$ is the distance in m from the
point in the source where neutrino emission occurs to the point in
the target where this neutrino is captured, and $E_{\nu}$ is the
neutrino energy in MeV.  In the two-zone experiment the source is
very compact, with typical linear dimension of 10--15~cm, and $L$ is
only about 1~m.  As a result the ripples of oscillation are strongly
manifested and are not averaged out when $\Delta m^2 \approx
1$~eV$^2$.

\item The density of gallium at its melting temperature of
29.8\degreesC is 6.095~g/cm$^3$ and the cross section for neutrino
capture on \nuc{71}{Ga} is $5.5 \times 10^{-45}$ cm$^2$.  These
factors ensure that the neutrino capture rate will be very high and
can be measured with good statistical accuracy.

\item The activity of the neutrino source can be measured in several
ways leading to a total uncertainty on the source emission rate as
low as 0.5\% (see, e.g., Ref.~\cite{ArPRC}).

\end{itemize}

In contrast, experiments with reactor and accelerator neutrinos
suffer from several disadvantages.  The neutrino energy $E_{\nu}$ is
distributed over a wide spectrum and the dimensions $L$ of the
sources and detectors are on the scale of several meters.  Other
disadvantages of a reactor or accelerator experiment are that the
knowledge of the neutrino flux incident on the target is usually
significantly worse than with a neutrino source and that, with some
targets, there are appreciable uncertainties in the cross section for
neutrino interaction.

To summarize, we propose a new experiment in which a very-intense
\nuc{51}{Cr} source irradiates a target of Ga metal that is divided
into two zones and the neutrino capture rate in each zone is
independently measured.  If there is either a significant difference
between the capture rates in the two zones, or the average rate in
both zones is considerably below the expected rate, then there is
evidence of nonstandard neutrino properties.  In the former case this
inference is independent of the nuclear physics uncertainty in the
detection cross section, but in the latter case it is important that
the cross section be known to within 5\%.

The proposed experiment has the potential to test neutrino
oscillation transitions with mass-squared difference $\Delta
m^2>0.5$~eV$^2$.  This capability exists because the experiment uses
a compact nearly-monochromatic neutrino source with well-known
activity, the dense target of Ga metal provides a high interaction
rate, and the special target geometry makes it possible to study the
dependence of the rate on the distance to the source.

Finally, we should point out that there are cosmological arguments
that preclude the existence of transitions to a sterile neutrino in
the mass-mixing angle region to which our experiment is sensitive
\cite{Dodelson06} \cite{Kusenko09}.  These arguments are, however,
not model independent and, in common with many others
\cite{other_sterile_exp}, we believe that direct measurements that
fully cover this region of parameter space are essential.

\begin{acknowledgments}

We are grateful to W. Haxton, H. Ejiri, M. Libanov, V. Matveev,
V. Rubakov, and A. Smirnov  for fruitful discussions.

\end{acknowledgments}

\end{document}